\documentclass{ws-procs975x65}

\begin{document}

\title{Monte Carlo simulation of the nuclear medium: Fermi gases, nuclei and the role of Pauli potentials.}

\author{M. \'ANGELES	 P\'EREZ-GARC\'IA}

\address{Department of Fundamental Physics,\\
 University of Salamanca, Spain \\ and \\
Instituto Universitario de F\'isica Fundamental y Matem\'aticas,\\
Facultad de  Ciencias, Plaza de la Merced s/n\\
Salamanca, E-37008, Spain\\
$^*$E-mail: mperezga@usal.es\\
www.usal.es}

\begin{abstract}
The role of Pauli potentials in the semiclassical simulation of Fermi gases at low temperatures 
is investigated. 
An alternative Pauli potential to the usual bivariate Gaussian form by Dorso {\it et al}~\cite{dorso} 
is proposed. This new Pauli potential allows for a simultaneous good reproduction of not  only the kinetic energy per particle but also the momentum distribution and 
the two-body correlation function. The reproduction of the binding energies in finite nuclei in the low and
medium mass range is also analyzed. What is found is that given a reasonable short-range 
atractive nuclear interaction one can include correlation effects in a suitable chosen 
density dependent Pauli potential.
\end{abstract}


\bodymatter

\section{Formalism}\label{aba:sec1}

Nuclear many-body simulations are a useful tool to study the relevant properties
of the nuclear medium in the thermodynamic conditions arising in matter in 
the aftermath of a Supernova event or in Neutron Stars. 
Examples of this are, for instance, nuclear {\it pastas}~\cite{pasta,maru} at 
densities in the range $0.01\rho_0 \le \rho \le 0.5 \rho_0$  ($\rho_0=0.148\,fm^{-3}$) 
and temperatures of decens of MeV or in heavy ion collisions~\cite{heavy}.
This type of simulations based on Monte Carlo or Molecular Dynamics techniques 
allow for a  dynamical description of the nuclear medium usually by using an effective
 interaction hamiltonian in a semiclassical treatment. 
In fermionic systems the genuine antisymmetrization
of the wave function is considered trough the inclusion of a Pauli potential. Pioneering
works on this line include those of Wilets{\it et al}~\cite{willets}. 
In this work the hamiltonian used to study the low temperature nucleon systems 
consists of a kinetic energy term and a Pauli effective potential ($V_{Pauli}$).

\begin{equation}
H=\sum_{i=1}^{N} \frac{{\bf p}_{i}^{2}}{2m_N}
+ \sum_{i=1,j>i}^{A} V_{Pauli} (r_{ij},p_{ij})\delta_{\tau_i \tau_j}
\delta_{\sigma_i \sigma_j},  
\label{ham}
\end{equation}
where $\delta_{\tau_i \tau_j}$ ($\delta_{\sigma_i \sigma_j}$) is  
the Kronecker's delta for the nucleon isospin (spin) third-component. 
${\bf p}_{i}$ is the 3-momentum of $i$-th nucleon and 
$r_{ij}=|{\bf r}_i-{\bf r}_j|$ ($p_{ij}=|{\bf p}_i-{\bf p}_j|$) 
the relative distance (momentum) of the $i$-th and $j$-th nucleons.

We will consider, for the sake of comparison, two ways to implement this potential.

\indent i) A Gaussian form introduced by Dorso {\it et al.}~\cite{dorso}, 

\begin{equation}
{V}_{Pauli}({r}_{ij}, {p}_{ij})= 
V_S\,\,  \exp \left(-\frac{r_{ij}^2}{2q_0^2}
-\frac{p_{ij}^2}{2p_0^2}\right), 
\label{Paulipot}
\end{equation}
Here $p_{0}$ and $q_{0}$ are momentum and length scales related to the
excluded phase-space volume that is used to mimic fermionic correlations and $V_S$ 
is the Pauli potential strength. All three parameters have been adjusted to reproduce
only the kinetic energy of a low temperature Fermi gas.

\indent ii) A new form proposed, based on spatial and momentum-dependent, 
two-body terms of the following form~\cite{jorge} 
\begin{equation}
{V}^{new}_{Pauli}({r}_{ij}, {p}_{ij})= 
V_{q}\exp(-r_{ij}/q_{0})+V_{p}\exp(-p_{ij}/p_{0})
+ V_{\Theta}\,\Theta_{\eta}(q_{i})\;, 
\label{VPauli1}
\end{equation}

where $q_{i}\!=\!|{\bf p}_{i}|/p_{\rm F}$,
and $\Theta_{\eta}$ is a smeared Heaviside-step function, 
$\Theta_{\eta}(q)\equiv \frac{1}{1+\exp[-\eta(q^{2}-1)]}$ and 
$\Theta_{\eta}(q)\longrightarrow\Theta(q)$ when $\eta$ is sufficiently big.
The parameters of the new Pauli potential $V_{q},V_{p},V_{\Theta}$ and $q_{0},p_{0},\eta$
will be adjusted to reproduce the kinetic energy per particle and both the momentum distribution and
two-body correlation function of a low-temperature Fermi gas.  The first and second terms
in the potential penalize two particles with the same quantum numbers coming together either in
space or momentum. This retains the essence of the fermionic wave function given by the 
Slater determinant. The third term forbids  any particle from having a momentum significantly
larger than the Fermi momentum. 

In both cases the potential parameters will depend on the density of the system and this 
will be crucial in
reproducing experimental binding energies when simulating low and medium mass nuclei as will be shown later.
The values for the parameters at saturation density $\rho_0$ and reduced
temperature $\tau=T/T_F=0.05$ are given in Table 1.

\begin{table}
\tbl{Pauli potential parameters.}
{\begin{tabular}{@{}cccc@{}}
\toprule
Pauli potential & Potential strength & $q_0$  & $p_0$ \\
& (MeV) & (fm) & (MeV/c) \\
\colrule
Dorso {\it et al} \cite{dorso} \hphantom{00} & \hphantom{0}$V_S=207$ & \hphantom{0}1.644 & 120\\
This work  \hphantom{00} & \hphantom{0}$V_q=13.517$, $Vp=1.260$, $V_{\Theta}=3.560$ ($\eta=30$)
 & \hphantom{0}0.66 & 49.03 \\
\botrule
\end{tabular}
}
\label{aba:tbl1}
\end{table}

\section{Results}\label{aba:sec2}

The simulations are performed  in a NVT system with $\tau=T/T_F$, and
 $N$ fermions in a cubic box of volume $V=L^3=N/\rho$. Then, using the Metropolis algorithm 
the system is thermalized until the stage where configurations are sampled 
in order to calculate the statistical averages for the magnitudes discussed below.

\begin{figure}
\begin{center}
\psfig{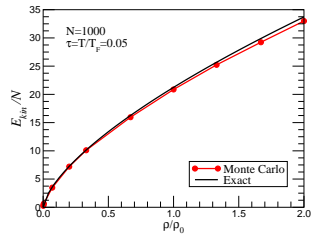}
\end{center}
\caption{Energy per particle of a Fermi gas simulated with $N=1000$ particles at 
$T=0.05 T_F$ as a function of density.}
\label{aba:fig1}
\end{figure}

In Fig.~\ref{aba:fig1} the red line shows the kinetic energy per particle  for a Fermi gas system with $N=1000$ particles
at $\tau=0.05$ as a function of density calculated using the new form of the Pauli 
potential Eq.(\ref{VPauli1}). Also plot with a black line is the exact result. 
We can see that there is a good reproduction of the kinetic energy.  

\begin{figure}[h]
\begin{center}
\psfig{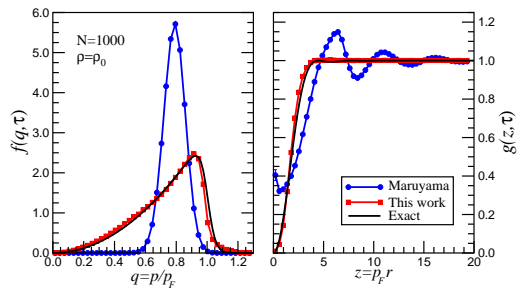}
\end{center}
\caption{Momentum distribution function and two-body correlation function of a Fermi gas. 
See text for details.}
\label{aba:fig2}
\end{figure}

In Fig.~\ref{aba:fig2} we can see on the left side the momentum distribution function $f(q,\tau)$ and, on the right
 side, the
two-body correlation function $g(z,\tau)$ with $z=p_Fr$. The blue curves correspond to the Dorso 
potential Eq.(\ref{Paulipot}) and the red curves to the new Pauli potential proposed 
Eq.(\ref{VPauli1}). Again the black line shows 
the exact result. We can see that a simultaneous good reproduction of both magnitudes is achieved with
the alternative new potential but not with the Dorso version. Particurlarly the "Fermi hole"  fails 
to be reproduced at small 
distances with the Dorso potential. This should be emphasized since these models are used in nuclear
 many-body simulations as in nuclear {\it pastas} as, for instance, in the work by Maruyama {\it 
et al}~\cite{maru}. The velocity distribution \cite{jorge}, not shown here, however peaks at lower values than the momentum
 distribution due to the fact that canonical and
kinematical momentum are not the same quantities~\cite{fai}. This is a genuine feature 
in this treatment with momentum dependent Pauli potentials in a hamiltonian formalism.

We now show finite nuclei simulation\cite{nuclei} results calculated with a simplified square-well 
nuclear potential with $V_{well}=-3$ MeV of width 2 fm and a core with $V_{core}=10$ MeV and width 1 fm.
 Coulomb interaction is also included. In Fig.~\ref{fig3.4}(a) binding energy per particle for a low to medium mass set of spin saturated 
symmetric nuclei of $A$ nucleons. As can be seen in Fig.~\ref{fig3.4}(b) kinetic(dashed line) 
and potential(dotted line) energy balance to obtain the total binding energy (solid line) per particle.
The density dependence of the parameters of the Pauli potential is crucial to provide enough positive 
contribution
to the linearly A-growing negative potential energy \cite{nuclei} and reproduce the experimental
 binding energy curve.

\def\figsubcap#1{\par\noindent\centering\footnotesize(#1)}
\begin{figure}[t]%
\begin{center}
  \parbox{2.1in}{\epsfig{figure=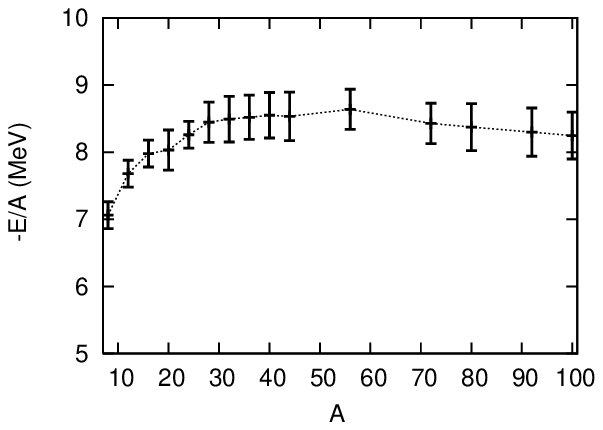,width=2in,scale=1.8}\figsubcap{a}}
  \hspace*{4pt}
  \parbox{2.1in}{\epsfig{figure=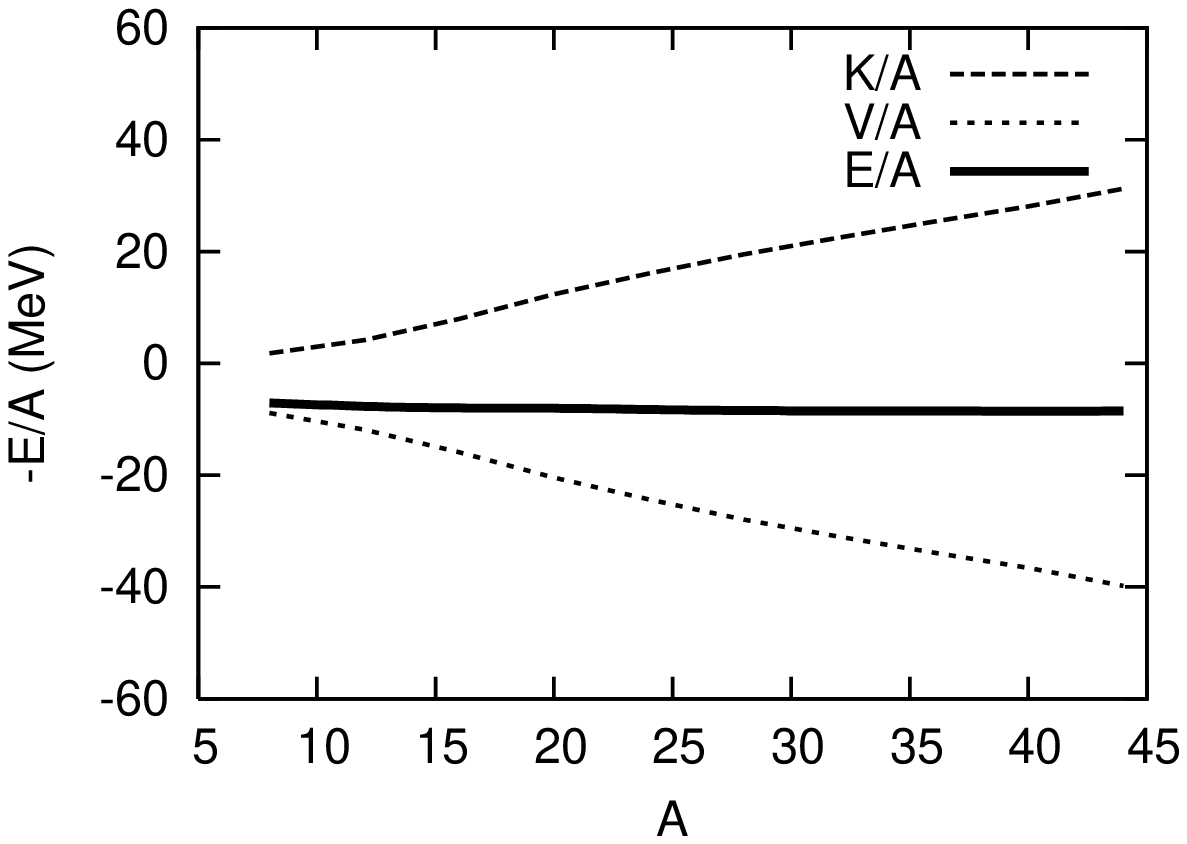,angle=-90,width=2in,scale=1.8}\figsubcap{b}}
  \caption{Application to nuclei. (a) Binding energy. (b) Kinetic and potential
 contributions to the binding energy.}%
  \label{fig3.4}
\end{center}
\end{figure}

\section*{Acknowledgments}
We acknowledge J. Piekarewicz, J. Taruna, K. Tsushima and A. Valcarce who are collaborators 
in this work. Partial funding has been provided by project DGI-FIS2006-05319.

\end{document}